\documentclass{aa}  
\usepackage{graphicx}
\usepackage{txfonts}
\usepackage{natbib}
\begin{document}
   \title{The mixed chemistry phenomenon in Galactic Bulge PNe\thanks{Based on 
observations made with the Spitzer Space Telescope, which is operated by the 
Jet Propulsion Laboratory, California Institute of Technology, under NASA 
contract 1407.} }
	  \titlerunning{Mixed chemistry in GB PNe}

   \author{J. V. Perea-Calder\'on\inst{1} \and 
           D. A. Garc\'{\i}a-Hern\'andez\inst{2} \and 
           P. Garc\'{\i}a-Lario\inst{3} \and 
           R. Szczerba\inst{4} \and 
           M. Bobrowsky\inst{5} 
}
	  \authorrunning{Perea-Calder\'on et al.}
  \offprints{J.V. Perea-Calder\'on}

  \institute{European Space Astronomy Centre, INSA S.A.  
     P.O. Box 78, E$-$28080 Madrid, Spain \email{Jose.Perea@sciops.esa.int}
  \and Instituto de Astrof\'{\i}sica de Canarias, C/ Via L\'actea s/n,
     38200 La Laguna, Spain  
  \and Herschel Science Centre. European Space Astronomy Centre, Research and
       Scientific Support Department of ESA, Villafranca del Castillo, 
       P.O. Box 78, E$-$28080 Madrid, Spain.
  \and N. Copernicus Astronomical Center, Rabia\'nska 8, 87-100 Toru\'n, Poland
  \and Department of Physics.
         University of Maryland, College Park, MD 20742-4111, USA
}

   \date{Received xxxx xx, 2008; accepted xxxx xx, 2009}
%
  \abstract
{} 
 {We investigate the dual-dust chemistry phenomenon in planetary nebulae (PNe)
and discuss reasons for its occurrence, by analyzing Spitzer/IRS spectra of a
sample of 40 Galactic PNe among which 26 belong to the Galactic Bulge (GB).}
 {The mixed chemistry is derived from the simultaneous detection of Polycyclic 
Aromatic Hydrocarbon (PAH) features in the 6$-$14\,$\mu$m range and crystalline
silicates beyond 20 $\mu$m in the Spitzer/IRS spectra.}
 {Out of the 26 planetary nebulae observed in the Galactic Bulge, 21 show
signatures of dual-dust chemistry. Our observations reveal that
the simultaneous presence of oxygen and carbon-rich dust features in the
infrared spectra of [WC]-type planetary nebulae is not restricted to late/cool 
[WC]-type stars, as previously suggested in the literature, but is a common
feature associated with all [WC]-type planetary nebulae. Surprisingly, we found 
that the dual-dust chemistry is seen also in all observed weak
emission-line stars ({\it wels}), as well as in other planetary nebulae with central stars 
being neither [WC] nor {\it wels}. Most sources observed display crystalline 
silicate features in their spectra, with only a few PNe
exhibiting, in addition, amorphous silicate bands.}
{We appear to detect a recent change of chemistry at the end of the 
Asymptotic Giant Branch (AGB)
evolution in the low-mass, high-metallicity population of GB PNe
observed. The deficit of C-rich AGB stars in this environment
suggests that the process of PAH formation in PNe occurs at 
the very end of the AGB phase.  In addition, the population of
low-mass, O-rich AGB stars in the Galactic Bulge, do not exhibit crystalline silicate features in 
their spectra. Thus, the high detection rate of dual-dust 
chemistry that we find cannot be explained by long-lived O-rich (primordial or 
circumbinary) disks. Our most plausible scenario is a final thermal 
pulse on the AGB (or just after), which could produce enhanced mass loss, 
capable of removing/mixing (sometimes completely) the remaining H-rich envelope and
exposing the internal C-rich layers, and generating shocks
responsible for the silicate crystallization.}

    \maketitle
%

\section{Introduction}

Planetary Nebulae (PNe) are the result of the evolution of low- and
intermediate-mass stars (1$-$8\,M$_{\odot}$), where the circumstellar gas
previously expelled during the Asymptotic Giant Branch (AGB) phase is ionized
by the central star (CS). The infrared emission from dusty circumstellar shell
that developes at the end of the AGB can remain strong during the PN phase, and thus
be easily detectable with infrared telescopes.

An interesting class of PNe are those with C-rich Wolf-Rayet ([WC]-type) central
stars (the so-called `WCPNe'), which have little or no H in their atmospheres. 
Observations completed by the Infrared Space Observatory (ISO) suggest
that the shells around some WCPNe exhibit a characteristic dual-dust  (or
mixed) chemistry \citep{Waters1998a, Waters1998b, Cohen1999, Cohen2002}, which
is not yet well understood. This is deduced from the simultaneous  presence of
the well-known family of infrared bands at 3.3, 6.2,
``7.7'', 8.6, and 11.3\,$\mu$m due to Policyclic Aromatic Hydrocarbons (PAHs: C-rich) 
\citep{Leger1984, Allamandola1985}, and several relatively broad  bands (the
strongest at 23.5, 27.5, and 33.8\,$\mu$m) attributed to crystalline  silicates
(O-rich), such as olivines and pyroxenes \citep{Waters1998a} in the  ISO
spectra. The mixed chemistry phenomenon appears to be common among  late-type/cool
WCPNe, while only one (NGC 5315) out of 6 early-type/hot WCPNe  observed with
ISO shows this dual-dust chemistry. In total  about 50\% of the WCPNe show
mixed chemistry \citep{Cohen2002,  DeMarco-Soker2002}. Note, however, that PAH
features seem to be present in  most of the 16 WCPNe analyzed by
\citet{Szczerba2001}, with the clear exception of K\,2-16. Therefore, we should
keep in mind that, due to the poor quality of ISO band 4
data, the true percentage of WCPNe with mixed chemistry may be underestimated.

In this {\it Letter}, we report the results of a systematic, 
spectroscopic analysis completed with the Infrared Spectrograph (IRS, \citealt{Houck2004})
onboard the Spitzer Space Telescope \citep{Werner2004} of a sample of 26
Galactic Bulge (GB) PNe around central stars, which are classified as
late-type/cool [WC], early-type/hot [WC], weak emission-line stars ({\it
wels}) \citep{Tylenda1993}, and neither [WC] nor {\it
wels}. We also briefly discuss the remaining 14 non-GB
PNe in our sample. Section 2 describes the infrared observations carried out and the data reduction
performed, while the results obtained are presented in Sect. 3. Finally, in
Sect.4 we discuss our findings in the context of stellar evolution in the
Galactic Bulge.

\begin{table*}
\caption{The sample of 21 Galactic Bulge PNe with dual-dust chemistry observed
 with Spitzer/IRS (program $\#$3633).}             
\label{GBdualdust}      
\begin{tabular}{r l c c c c c c}		
\hline 
\hline 
IRAS name    & PN name  & R.A.       & DEC       &  Spectral type$^{\mathrm{a}}$         & Optical diameter$^{\mathrm{d}}$ & distance from GC & dust type \\ 
             &          & J2000.0    & J2000.0   &                        & (arcsec)                        & ($\degr$)        &           \\
\hline
17480$-$3023 & M 3-44   & 17 51 18.7 & -30 23 53 & [WC11]$^{\mathrm{b}}$  & 4.4 & 1.91 & mixed \\
17549$-$3347 & H 1-43   & 17 58 14.5 & -33 47 37 & [WC11]$^{\mathrm{b}}$  & 2.0 & 5.56 & mixed \\
17434$-$3307 & M 1-27   & 17 46 45.5 & -33 08 35 & [WC11]$^{\mathrm{b}}$ & 5.3 & 4.21 & mixed \\
18061$-$2505 & MaC 1-10 & 18 09 12.4 & -25 04 35 & [WC8]$^{\mathrm{b}}$   &10.0 & 6.52 & mixed \\
17355$-$2206 & M 1-25   & 17 38 30.3 & -22 08 39 & [WC6]$^{\mathrm{b}}$   & 4.6 & 6.99 & mixed \\ 
17425$-$2056 & M 3-15   & 17 45 31.8 & -20 58 06 & [WC5]$^{\mathrm{b}}$   & 4.2 & 7.97 & mixed \\
17388$-$2440 & Hb 4     & 17 41 53.1 & -24 42 03 & [WC4]$^{\mathrm{b}}$   & 6.2 & 4.32 & mixed \\ 
18259$-$3132 & Cn 1-5   & 18 29 11.7 & -31 29 59 & [WC4]$^{\mathrm{b}}$   & 7.0 & 9.75 & mixed \\
17389$-$2409 & M 2-14   & 17 41 57.7 & -24 11 11 & {\it wels}$^{\mathrm{b}}$    & 0.0 & 4.82 & mixed \\ 
17496$-$2221 & M 1-31   & 17 52 41.4 & -22 21 57 & {\it wels}$^{\mathrm{b}}$    & 0.0 & 6.76 & mixed \\
18006$-$3117 & M 2-27   & 18 03 52.6 & -31 17 47 & {\it wels}$^{\mathrm{b}}$    & 4.8 & 4.60 & mixed \\
18054$-$2217 & M 1-40   & 18 08 26.0 & -22 16 53 & {\it wels}$^{\mathrm{b}}$    & 5.0 & 8.40 & mixed \\
18094$-$2450 & H 1-61   & 18 12 34.0 & -24 50 01 & {\it wels}$^{\mathrm{b}}$    & 0.0 & 7.27 & mixed \\
17156$-$3135 & Th 3-4   & 17 18 51.9 & -31 39 06 & none$^{\mathrm{b}}$    & 0.0 & 6.38 & mixed \\
17178$-$2900 & M 3-38   & 17 21 04.3 & -29 02 60 & none$^{\mathrm{c}}$    & 0.0 & 5.37 & mixed \\
17217$-$2803 & M 3-8    & 17 24 52.2 & -28 05 54 & none$^{\mathrm{b}}$    & 5.4 & 4.64 & mixed$^{\mathrm{e}}$ \\
17262$-$2623 & H 1-16   & 17 29 23.4 & -26 26 05 & none$^{\mathrm{b}}$    & $<$5.0 & 4.37 & mixed \\
17521$-$2144 & Hb 6     & 17 55 07.1 & -21 44 41 & none$^{\mathrm{b}}$    & 5.0 & 7.50 & mixed \\
17523$-$3033 & H 1-40   & 17 55 36.0 & -30 33 32 & none$^{\mathrm{b}}$    & 3.8 & 2.71 & mixed \\
18006$-$3241 & H 1-50   & 18 03 53.5 & -32 41 42 & none$^{\mathrm{b}}$    & $<$10. & 5.43 & mixed \\
18100$-$3220 & H 1-62   & 18 13 17.9 & -32 19 42 & unknown                & 0.0 & 6.85 &  mixed \\
\hline
\hline                  
\end{tabular}
\smallskip
\smallskip

\begin{tabular}{p{\textwidth}}
$^{\mathrm{a}}$ none - neither [WC]~nor {\it wels};
$^{\mathrm{b}}$ \cite{Gorny2004}; 
$^{\mathrm{c}}$ \cite{Gorny2008};
$^{\mathrm{d}}$ Strasbourg-ESO Catalogue of Galactic Planetary Nebulae, \\
\cite{Acker1992}, 0.0 - PN is a point-like 
source; $^{\mathrm{e}}$ PAHs presence is tentative.
\end{tabular}
\end{table*}

\section{Observations and data reduction}

The spectral data were taken with Spitzer/IRS under a General Observer
program ($\#$3633, PI, M. Bobrowsky). We obtained Spitzer/IRS spectra of a
sample of 40 PNe\footnote{Plots of all spectra are available online. Table\,1
and online Tables\,A.1 and A.2 collect relevant information for our sample of
PNe.} in the 5.2$-$37.2 $\mu$m range by using the Short-Low (SL: 5.2$-$14.5
$\mu$m; 64$<$R$<$128), Short-High (SH: 9.9$-$19.6 $\mu$m; R$\sim$600) and
Long-High (LH: 18.7$-$37.2 $\mu$m; R$\sim$600) modules. All sources in our
sample are bright at mid-infrared wavelengths, and it was easy to achieve a
signal-to-noise ratio higher than 50 in each of the three modules SL, SH, and LH by using two
exposure cycles of 6 s. The starting point for our interactive data reduction
were the coadded 2D flat-fielded images (one for each nod position; pipeline
version 12.3), in which ``rogue'' pixels were cleaned.
For SL, the two nod positions were subtracted in order
to remove the sky background. However,
for the high-resolution modules no background subtraction was done since no sky
measurements were taken and the SH and LH slits were too small for on-slit
background subtraction. The spectra for each nod position were extracted from
the 2D images, and wavelength and flux calibrated using the Spitzer IRS Custom
Extractor (SPICE) with a point-source aperture. The 1D spectra were cleaned for
bad data points, spurious jumps and glitches, combined and merged into one
single spectrum per module for each target using SMART\footnote{SMART was
developed by the IRS Team at Cornell University and is available through the
Spitzer Science Center at Caltech.} \citep{Higdon2004}.
Finally, when required we scaled the LH spectra to the available Midcourse Space Experiment
\citep[MSX;][]{Egan_etal_2003} photometry at
21.34\,$\mu$m, and the corresponding SL and SH spectra were also scaled upward to match the LH spectra.
This is because some of the PNe studied here are extended sources (optical
diameter $>$4"), giving a flux loss at the smaller SL and SH apertures. The
MSX photometry at 21.34 $\mu$m was selected as a reference point for flux 
calibration because this filter is less affected by 
nebular emission lines and dust emission features.

\vspace{-0.09in} 
\section{Results}
In Table\,1, we present the names (columns 1 \& 2), coordinates (columns 3 \& 4), 
spectral type (column 5), optical diameter (column 6), distance from Galactic 
Center (GC: column 7), and dust type (column 8) for the 21 PNe in our
sample, which belong to the Galactic Bulge\footnote{A PN was counted as a member
of the GB when its angular distance from the GC was $<$\,10$\degr$ and its
optical angular diameter was $<$\,10$\arcsec$.} and show dual-dust
chemistry. The total number of GB PNe observed by us is 26, so the 
occurrence of dual-dust chemistry (21/26) is much higher than the 6/11
observed in the sample of GB PNe discussed by \citet{Gutenkunst2008}.
   \begin{figure}
   \centering
   \includegraphics[bb=0 22 435 770,width=8cm,clip]{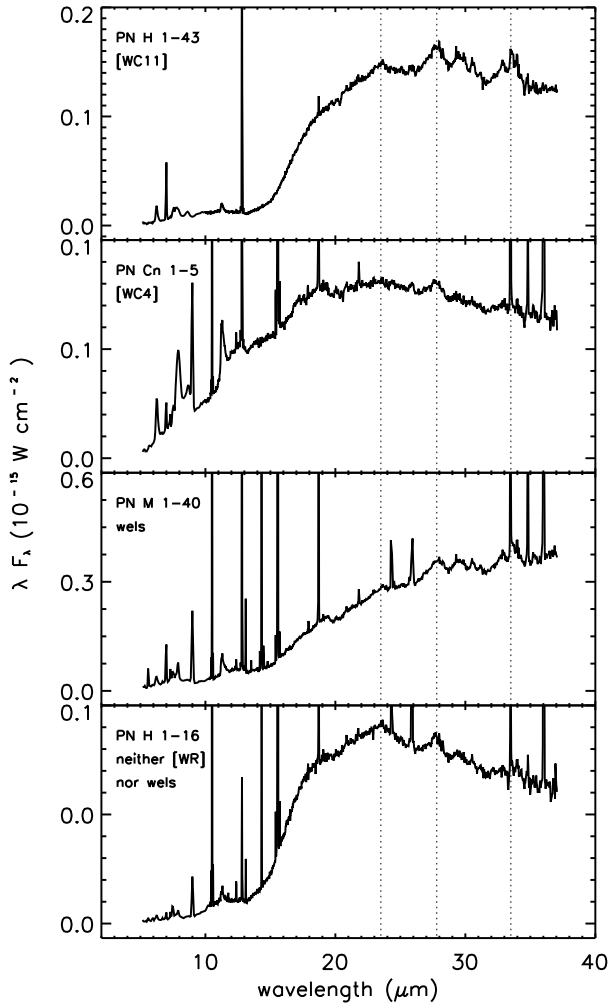}
   \caption{Spitzer/IRS spectra of four representative GB PNe from our sample 
(from top to bottom: late-type/cool [WC], early-type/hot [WC], {\it wels} and 
neither [WC] nor {\it wels}). Note the simultaneous detection of PAH emission 
features in the 6$-$14 $\mu$m range (see Fig.\,2) together with prominent 
crystalline silicate emission features (marked with dotted vertical lines at 
23.5, 27.5 and 33.8\,$\mu$m) beyond 20 $\mu$m, in all cases.} 
   \label{GB4whole}
   \end{figure}
   \begin{figure}
   \centering
   \includegraphics[bb=0 22 435 770,width=8cm,clip]{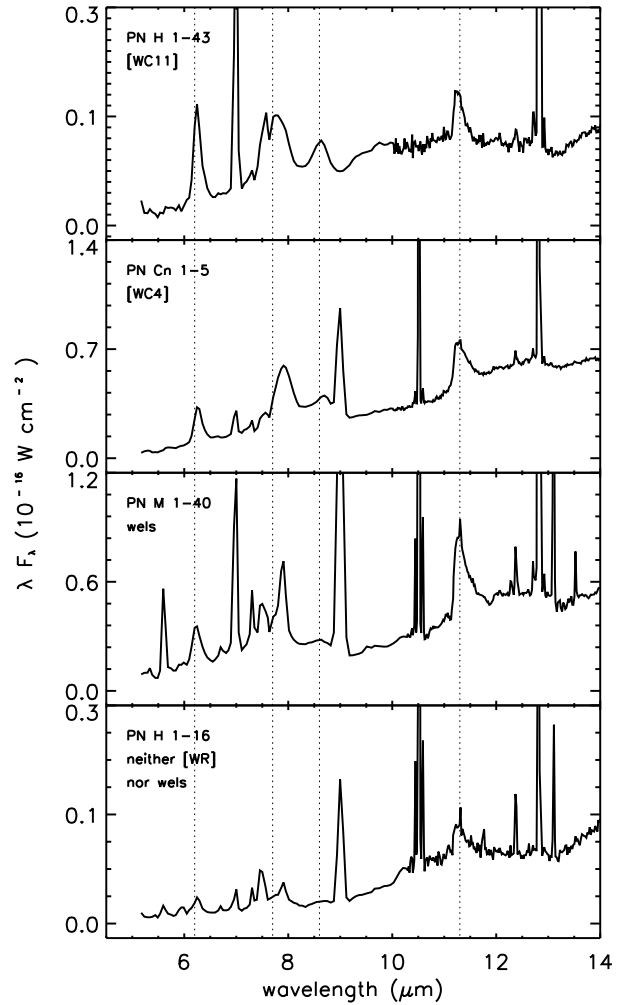}
   \caption{Same spectra as in Fig.\,1, but for wavelenghts $<$ 14\,$\mu$m. 
The positions of typical PAH features at 6.2, ``7.7'', 8.6 and 11.3 $\mu$m are 
indicated by dotted vertical lines.}
   \label{GB4zoom}
   \end{figure}
Figure\,1 shows representative Spitzer/IRS spectra corresponding to 4 classes of
PN CSs from Table\,1: late-type/cool [WC] (H 1-43); early-type/hot [WC] (Cn
1-5); {\it wels} (M 1-40); and neither [WC] nor {\it wels} (H 1-16). In Appendix
B, we present spectra for all 21 PNe from Table\,1 (see Figs.\,B1, B2 and B3),
and for the remaining 5 GB PNe in our sample (Table\,A1 in Appendix A
contains relevant information for these sources), which show only the
presence of silicates (Fig.\,B4). The spectral type of the CS is shown below the
PN name and  ``none'' is used instead of ``neither [WC] nor {\it wels}''. All
spectra exhibit a strong infrared continuum emission, which peaks  between 20 and
40\,$\mu$m, as observed in many other PNe detected by ISO. This strong
infrared excess is produced by the thermal emission of dust still present in
their remnant circumstellar AGB shells. Overimposed on this infrared continuum
we can clearly distinguish the presence of narrow nebular emission lines.

In the spectra, we can also see broad solid-state features from the dust 
grains. The simultaneous presence of PAH emission features (C-rich) in the 
6$-$14\,$\mu$m range (Fig.\,2 and the right panels of
Figs.\,B1, B2, and B3) and prominent crystalline silicate emission features (O-rich) beyond 20 $\mu$m
(Fig.\,1 and the left panels of Figs.\,B1, B2 and B3)
illustrates the dual-dust chemistry in all 21 GB PNe. The mixed chemistry is seen not only in
late-type, but also in all observed early\footnote{Note that Cn 1-5 was 
previously observed with ISO, although the spectrum was of bad quality and the
presence of dual-dust chemistry was uncertain in this early-type
WCPN.}-type WCPNe. However, the most striking result is that dual-dust chemistry
was detected in {\it all} observed {\it wels} as well as in some other GB
PNe with CSs, which are very likely of none of the above types as
inferred from their optical  spectra (see Table\,1). Spectral type of H 1-62 is unknown. 
The class of GB PNe with neither [WC] nor
{\it wels} CSs is investigated in a subsequent paper
(G\'orny et al.~2009, in preparation).

A broad, dust feature around 10\,$\mu$m is seen in a few PNe. This feature 
appears to be due to amorphous silicates and is seen only in 4 (Th 3-4, M 3-38, M
3-8 and H 1-40 - see Figs.\,B2 and B3) out of the 21 GB PNe with dual-dust
chemistry. The only other known case of WCPN with such feature is SwSt\,1
\citep[see e.g.][]{Szczerba2001}. By interpreting the strength of the 11.3\,$\mu$m PAH
band as a measure of the PAH abundance, we see that these 4 sources (except M
3-38) have the lowest PAH content (detection of PAHs is even tentative in 
the case of M 3-8 - see Table\,1 and Fig\,B3). Surprisingly, this 10\,$\mu$m
band is seen in most non dual-dust chemistry GB sources in our
sample, except H 1-12, for which we have only SH observations and we cannot
deduce the presence of this band, and M 3-13 whose
spectrum is of rather low quality (see Fig.\,B.4). These sources show only weak crystalline silicate
features. The sources with 10\,$\mu$m dust feature are
discussed in more detail by G\'orny et al.\,(2009, in preparation).

There are 14 additional PNe in our sample, which are not counted as GB sources. 
The relevant information for these sources is available only online and is 
collected in Table\,A2 with spectra shown in Figs.\,B5
and B6. Among them, there are 6 PNe with dual-dust chemistry (3 early-[WC]; 2 {\it wels} and one with 
unknown type CS). Surprisingly, there is one {\it wels}\footnote{The three 
{\it wels} PNe are located only slightly further than 10$\degr$ from the 
Galactic Centre, so they may as well belong to the GB population.} (M\,1-20)
that does not display the signs of O-rich dust, but only PAHs and a broad plateau emission 
between 11 and 14\,$\mu$m (see Fig.\,B5), attributed to primordial C-rich 
material, also observed in proto-PNe \citep[e.g.][]{Kwok2001}. This PN, 
and another C-rich source (Tc 1) showing unusual, hydrogenated PAHs (see Fig.\,B6)
will be discussed in more detail in a forthcoming paper. 
We note the exceptional source GLMP\,698 (with CS 
classified as [WC8]), for which only weak bands of crystalline silicates are 
detected (similar to ISO target K\,2-16 with a CS classified as [WC\,11]).
Another 5 sources show only silicates, with no indication of C-rich dust. 
Out of the 14 non-GB PNe, there are 5 sources showing both, amorphous and 
crystalline silicates. 

\section{Discussion}

The origin of the dual-dust chemistry in PNe is still unclear. In general, 
several mechanisms may play a role:

{\bf i}) a  very late thermal pulse at the end of the AGB phase (or  just
after), which may turn an O-rich outflow into a C-rich one \citep{Waters1998a}.
Due to the enhanced mass loss, high temperature annealing may lead to
the crystallization of amorphous silicates  \citep[e.g.][]{Sylvester1999}, while
PAHs form from the newly released C-rich gas;

{\bf ii}) evaporation of Oort-like cloud comets may release  crystalline
silicates \citep{Cohen1999}, while PAHs form after transformation of
the star into C-rich one;

{\bf iii}) the central star of the PN may exist in a binary system  with a
circumbinary O-rich disk that was formed long before the PN. Low temperature
crystallization may take place in such a disk
\citep{Molster1999}. PAHs would form after transformation of the O-rich star 
into a C-rich one;

{\bf iv}) a low-mass main sequence star, brown dwarf, or
planet may spiral into the
AGB star, inducing a chemistry change as a consequence of extra-mixing
\citep{DeMarco-Soker2002}. High temperature annealing of amorphous silicates
and PAH formation from the newly released C-rich gas could be invoked; 

{\bf v}) for intermediate-mass stars, the strong mass loss may induce a
sudden  deactivation of the {\it hot bottom burning}  \citep[see
e.g.][]{Gar-Her2006}  just at the end of the AGB evolution. The high
temperature annealing may lead to crystallization of amorphous silicates, while
carbon, again allowed to be dredged-up to the surface, may be used for the
PAH formation.

In our sample of GB PNe, we see an overwhelming presence of crystalline 
silicates \citep[see also][]{Gutenkunst2008}. This might be explained by the 
existence of long-lived O-rich disk (primordial or circumbinary) around CSs of 
PNe as proposed in scenarios {\bf ii)} and {\bf iii)}.
However, these explanations appear to be inconsistent with the existing observations of AGB stars in the GB,
which in general do not show any evidence of crystalline silicates 
\citep{Vanhollebeke2007}. In addition, \citet{Miszalski2008} 
reported that out of 300 GB PNe only some dozen (about
10-15\%) display signatures of periodic variability (interpreted as due to binarity) in the 
OGLE-III data. The lack of evidence for a dominant
binarity also excludes more exotic mechanisms, such as
these proposed in scenario {\bf iv}).
Scenario {\bf v}) is also not applicable to the GB PNe,
since the current AGB population in the Galactic Bulge
consists of mostly low-mass stars \citep[see e.g.][and references therein]{Uttenthaler_etal_2007}. 

To elucidate the validity of  the scenario {\bf i}), it is important to
take into account the higher (on average) metallicity of the GB population 
compared with the rest of Galaxy \citep{Chiappini2008}. Low-mass O-rich AGB 
stars in the GB, with limited efficiency in the third dredge-up, do not succeed 
in bringing enough carbon into the envelope to produce C-rich AGB stars in a 
high-metallicity environment. Hence, it seems very likely that the formation of C-rich 
PNe (so numerous in the GB) must originate in a final He-shell instabillity (probably
violent), which exposes the He- and C-rich layers only when the AGB phase is 
terminated.

The large mass loss expected in GB PNe at the end of the
AGB due to the higher metallicity compared with the average Galactic metallicity 
(and thus higher amount of dust) may facilitate the mixing/removal of the remaining hydrogen from
the stellar envelope leading to the formation of [WC] CSs. In addition,  this
violent process may be connected to the operation of strong shocks, which  can
increase the temperature in circumstellar shells leading to crystallization  of
silicates. Therefore, we suggest that process {\bf i}) proposed above appear 
to be the most plausible scenario in explaining the
formation of C-rich PNe with mixed  chemistry in the GB.

The detection of dual-dust chemistry in the majority of the GB PNe
constitutes the first evidence that this phenomenon is not restricted to
late-type WCPNe, as previously suggested. We suggest that the
dual-dust chemistry in the GB PNe is related to the low
mass and high metallicity of their precursors. The missing population of C-rich AGB stars in
the GB and the overall lack of crystalline silicates around GB AGB stars,
suggest that the formation of PNe with dual-chemistry
may be due to a final He-shell instability that leads to
a simultaneous change in chemistry (PAH formation) and
crystallization of  pre-existing silicates. However, other possibilities for dual-dust 
chemistry cannot be excluded in other than GB environment. The problem clearly deserves 
further theoretical analysis and needs to take into account the new observational
constraints provided here.

\begin{acknowledgements}
We acknowledge support from the Faculty of the European Space Astronomy Centre
(ESAC) and from the Comunidad de Madrid PRICIT project S-0505/ESP-0237
(ASTROCAM). RSz acknowledges support from grants N203 393334 and N203 0661 33.
We thank to Slawek G\'orny for his useful comments and help during the
preparation of this manuscript. We are also grateful to the referee, Dr. M.
Busso, whose comments allowed us to improve the manuscript.
\end{acknowledgements}

\Online
\appendix
\section{Tables}
%
\begin{table*}
\caption{The other Galactic Bulge PNe observed with Spitzer/IRS (program $\#$3633).}             
\label{GBother}      
\begin{tabular}{r l c c c c c c}		
\hline 
\hline 
IRAS name    & PN name  & R.A.       & DEC       &  Spectral type$^{\mathrm{a}}$         & Optical diameter$^{\mathrm{e}}$ & distance from GC & dust type \\ 
             &          & J2000.0    & J2000.0   &                        & (arcsec)                        & ($\degr$)        &           \\
\hline

17230$-$3459   & H 1-12   & 17 26 24.3 & -35 01 41 & unknown                & 6.8 & 7.33 & O \\
17385$-$2211   & M 3-13   & 17 41 36.6 & -22 13 03 & none$^{\mathrm{b}}$    & 0.0 & 6.78 & O \\
17427$-$3402   & H 1-32   & 17 46 06.3 & -34 03 45 & none$^{\mathrm{b}}$    & $<$5.0 & 5.13 & O \\
17459$-$3421   & H 1-35   & 17 49 14.0 & -34 22 52 & none$^{\mathrm{c}}$    & 2.0 & 5.50 & O \\
17585$-$2825   & M 2-23   & 18 01 42.7 & -28 25 44 & none$^{\mathrm{d}}$    & 8.5 & 3.56 & O \\
\hline
\hline                  
\end{tabular}
\smallskip
\smallskip

\begin{tabular}{p{\textwidth}}
$^{\mathrm{a}}$ none - neither [WC]~nor {\it wels};
$^{\mathrm{b}}$ G\'orny et al. (2004);
$^{\mathrm{c}}$ Wang \& Liu (2007);
$^{\mathrm{d}}$  G\'orny et al. (2008);
$^{\mathrm{e}}$ Strasbourg-ESO \\
Catalogue of Galactic Planetary Nebulae (Acker et al.\,1992),0.0 - PN is a point-like source
\end{tabular}
\end{table*}
%
\begin{table*}
\caption{The rest of PNe (non-Galactic Bulge) observed within Spitzer/IRS program $\#$3633.}             
\label{nonGBother}      
\begin{tabular}{r l c c c c c c}		
\hline 
\hline 
IRAS name    & PN name  & R.A.       & DEC       &  Spectral type$^{\mathrm{a}}$        & Optical diameter$^{\mathrm{f}}$& distance from GC & dust type \\ 
             &          & J2000.0    & J2000.0   &                        & (arcsec)                        & ($\degr$)        &           \\
\hline
17597$-$1442 & GLMP 698 & 18 02 38.3 & -14 42 03 & [WC8]$^{\mathrm{b}}$   & $--$& 14.77& O  \\
18307$-$1109 & M 1-51   & 18 33 29.2 & -11 07 26 & [WC4-6]$^{\mathrm{c}}$ & 9.5 & 21.03& mixed \\
17534$-$1628 & M 1-32   & 17 56 20.3 & -16 29 08 & [WC4]$^{\mathrm{d}}$   & 7.6 & 12.70 & mixed \\
18408$-$1347 & M 1-60   & 18 43 38.4 & -13 44 56 & [WC4]$^{\mathrm{d}}$   & $<$10.0 & 20.30 & mixed  \\ 
17260$-$1913 & M 1-20   & 17 28 57.6 & -19 15 53 & {\it wels}$^{\mathrm{e}}$ & $<$7.0 & 10.40 & C \\
18295$-$2510 & NGC 6644 & 18 32 34.7 & -25 07 44 & {\it wels}$^{\mathrm{d}}$    & 2.6 & 11.14& mixed \\  
18425$-$3323 & IC 4776  & 18 45 50.7 & -33 20 34 & {\it wels}$^{\mathrm{d}}$    & 7.5 & 13.60 & mixed \\ 
17561$-$1532 & M1-33    & 17 58 59.2 & -15 32 20 & unknown & 0.0 & 13.76 & mixed \\ 
17119$-$5926 & Hen 3-1357 & 17 16 21.1 & -59 29 23 & unknown & $--$ & 31.16 & O \\ 
17225$-$4408 & Cn 1-3   & 17 26 12.3 & -44 11 26 & unknown & $<5.0$ & 15.75 & O \\
17360$-$1815 & Hen 2-260& 17 38 57.4 & -18 17 36 & none$^{\mathrm{d}}$ & $<$10& 10.77 & O \\
17358$-$2854 & 19W32    & 17 39 02.9 & -28 56 35 & unknown & 24.0 & 1.44 & O \\
18308$-$2241 & IC 4732  & 18 33 54.6 & -22 38 40 & unknown & 4.0 & 12.57 & O \\
17418$-$4604 & Tc 1     & 17 45 35.3 & -46 05 23 & unknown & 9.6 & 17.20 & C \\ 
\hline
\hline                  
\end{tabular}
\smallskip
\smallskip

\begin{tabular}{p{\textwidth}}
$^{\mathrm{a}}$ none - neither [WC]~nor {\it wels};
$^{\mathrm{b}}$ Parker \& Morgan (2003)
$^{\mathrm{c}}$ Tylenda et al.\,(1993), G\'orny et al.\,(2001);
$^{\mathrm{d}}$ G\'orny et al.\,(2004);\\
$^{\mathrm{e}}$ G\'orny et al.\,(2008);
$^{\mathrm{f}}$ Strasbourg-ESO Catalogue of Galactic Planetary Nebulae (Acker et al.\,1992), 
0.0 - PN is a point-like \\
source, $--$ PN is not listed in the catalogue.
\end{tabular}
\end{table*}

\section{Spectra}
   \begin{figure*}
   \centering
   \includegraphics[width=\textwidth]{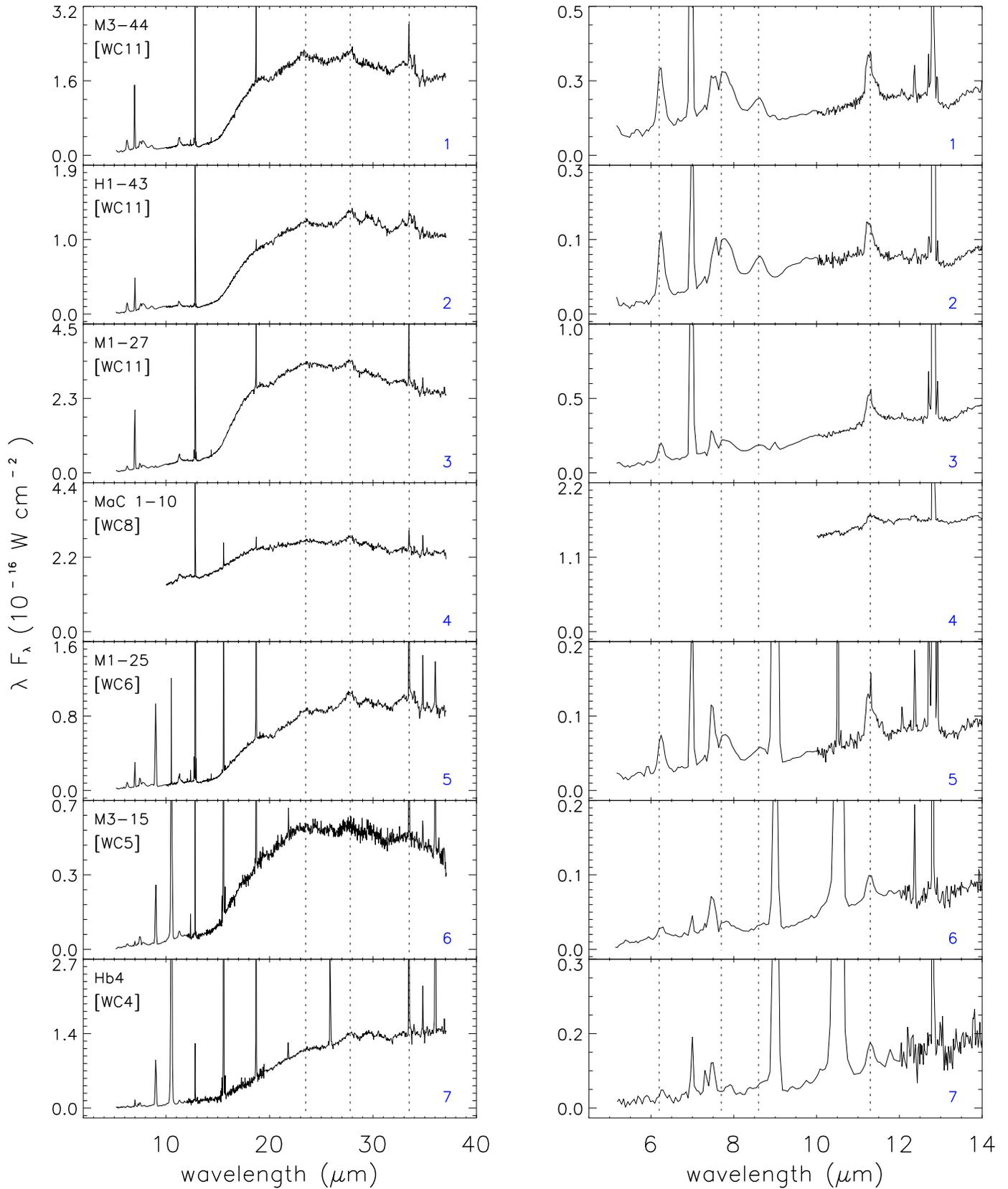}
   \caption{Spitzer/IRS spectra of Galactic Bulge PNe, which show dual-dust
chemistry (see Tabl\,1). The spectral type is shown under name of PN.
Note the simultaneous detection of PAH emission 
features in the 6$-$14 $\mu$m range (dashed vertical lines at 6.2, ``7.7'', 
8.6 and 11.3\,$\mu$m - right panels) together with crystalline silicate features 
(dashed vertical lines at 23.5, 27.5 and 33.8\,$\mu$m - left panels).}
   \label{spectraGB1}
   \end{figure*}
   \begin{figure*}
   \centering
   \includegraphics[width=\textwidth]{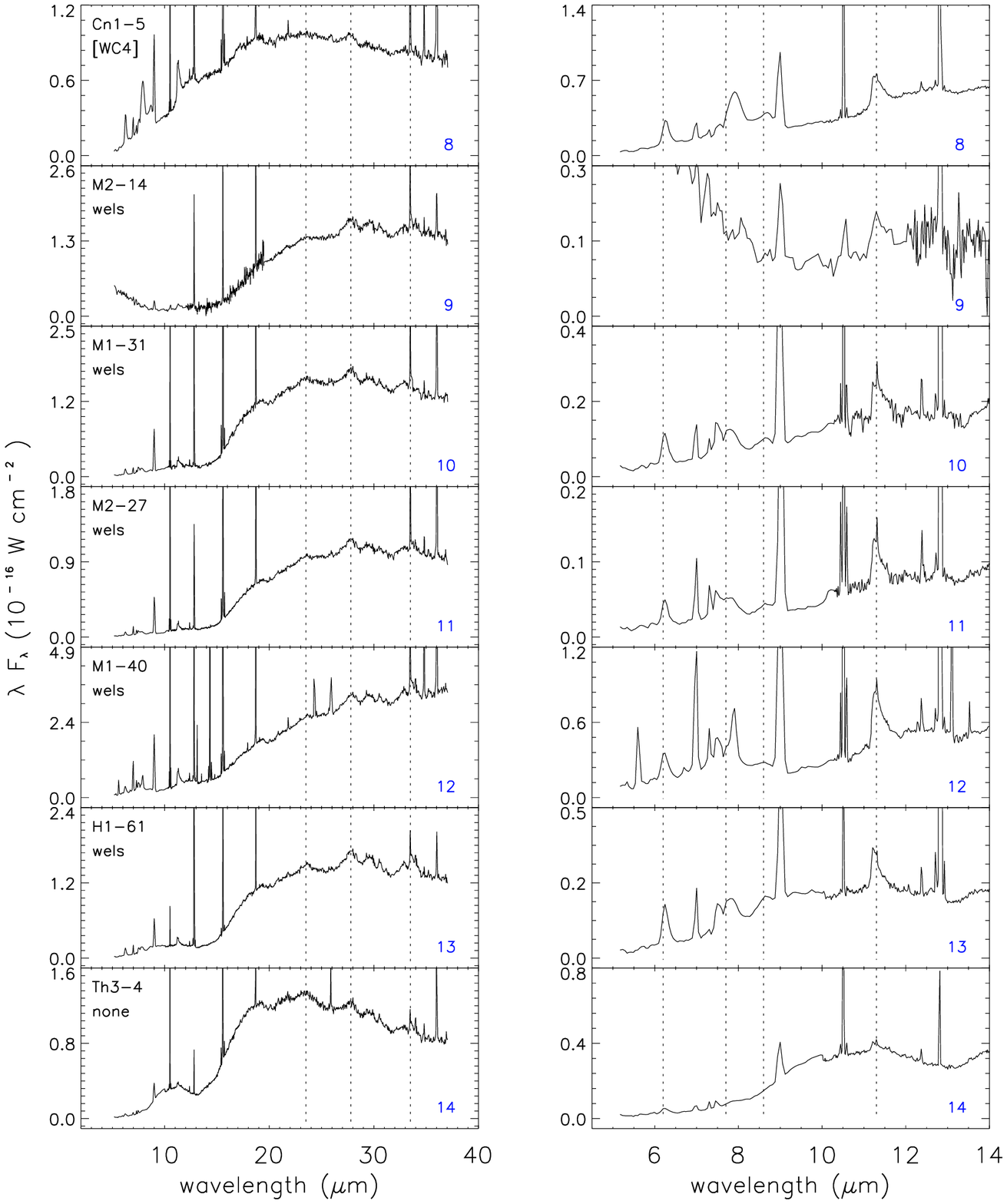}
   \caption{Spitzer/IRS spectra of Galactic Bulge PNe, which show dual-dust
chemistry {\it cont.} (see Tabl\,1). The spectral type is shown under name of PN; none --
means neither [WC] nor {\it wels}. Note the simultaneous detection of PAH 
emission features in the 6$-$14 $\mu$m range (dashed vertical lines at 6.2, 
``7.7'', 8.6 and 11.3\,$\mu$m - right panels) together with crystalline silicate 
features (dashed vertical lines at 23.5, 27.5 and 33.8\,$\mu$m - left panels).}
   \label{spectraGB2}
   \end{figure*}
   \begin{figure*}
   \centering
   \includegraphics[width=\textwidth]{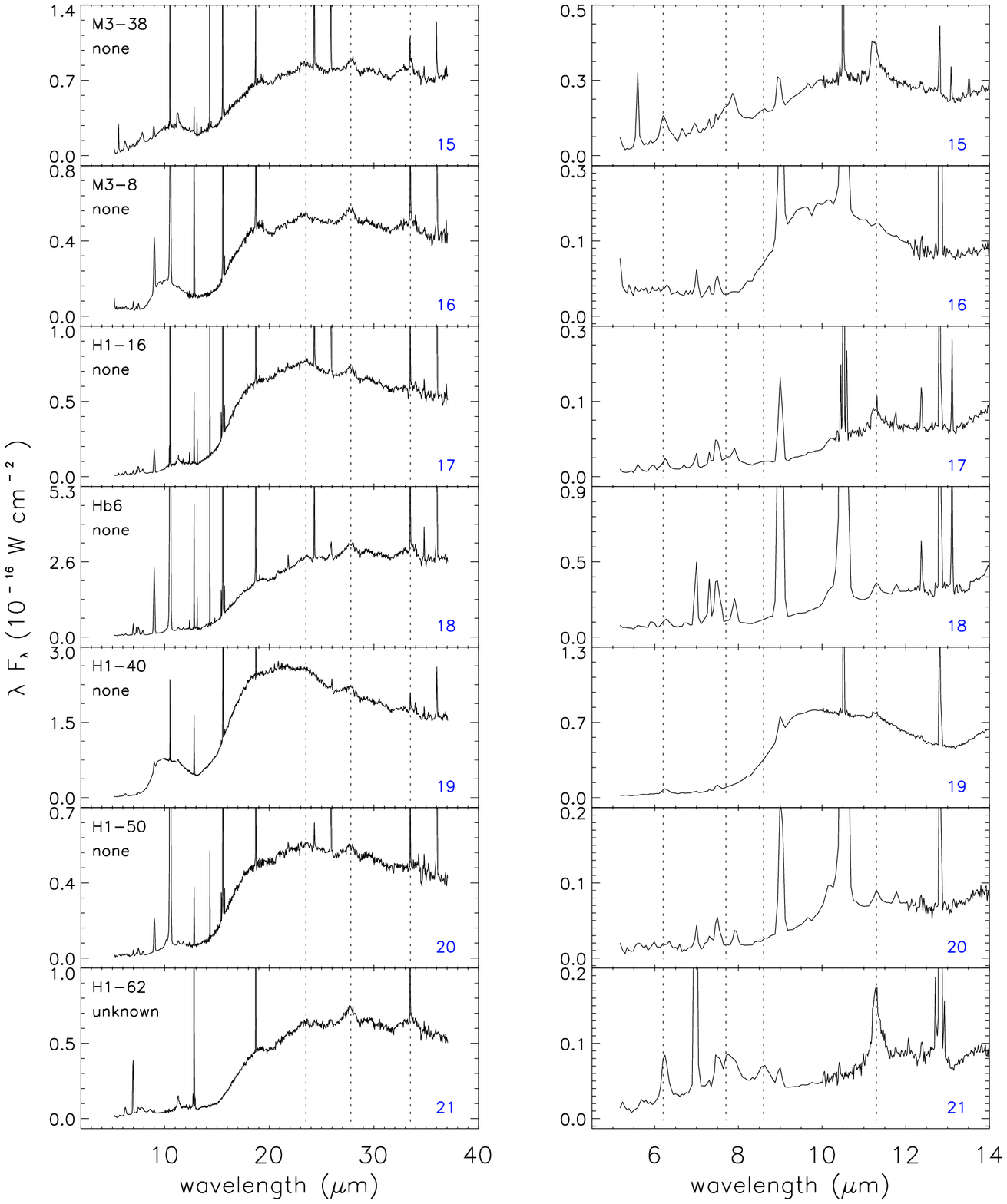}
   \caption{Spitzer/IRS spectra of Galactic Bulge PNe, which show dual-dust
chemistry {\it cont.} (see Tabl\,1). The spectral type is shown under name of PN; none --
means neither [WC] nor {\it wels}. Note the simultaneous detection of PAH 
emission features in the 6$-$14 $\mu$m range (dashed vertical lines at 6.2, 
``7.7'', 8.6 and 11.3\,$\mu$m - right panels) together with crystalline silicate 
features (dashed vertical lines at 23.5, 27.5 and 33.8\,$\mu$m - left panels).}
   \label{spectraGB3}
   \end{figure*}
   \begin{figure*}
   \centering
   \includegraphics[width=\textwidth]{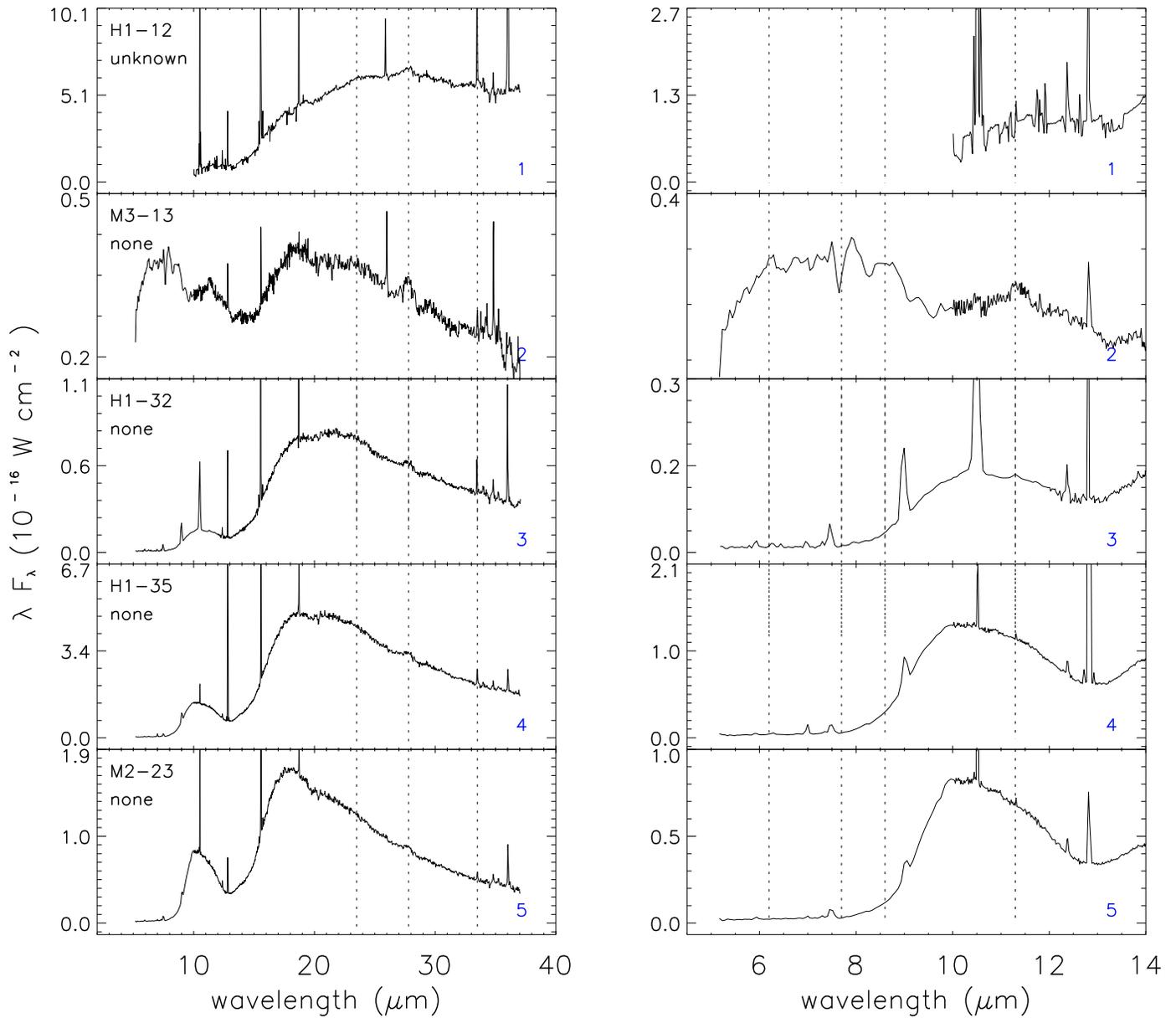}
   \caption{Spitzer/IRS spectra of Galactic Bulge PNe, which do not show dual-dust
chemistry (see Tabl\,A.1). The spectral type is shown under name of PN; none --
means neither [WC] nor {\it wels}. Dashed vertical lines at 23.5, 27.5 and 33.8\,$\mu$m
on the left panels mark position of crystalline silicate bands, while those
at 6.2, ``7.7'', 8.6 and 11.3\,$\mu$m on the right panels show position of PAH 
features.}
   \label{spectraGB4}
   \end{figure*}
   \begin{figure*}
   \centering
   \includegraphics[width=\textwidth]{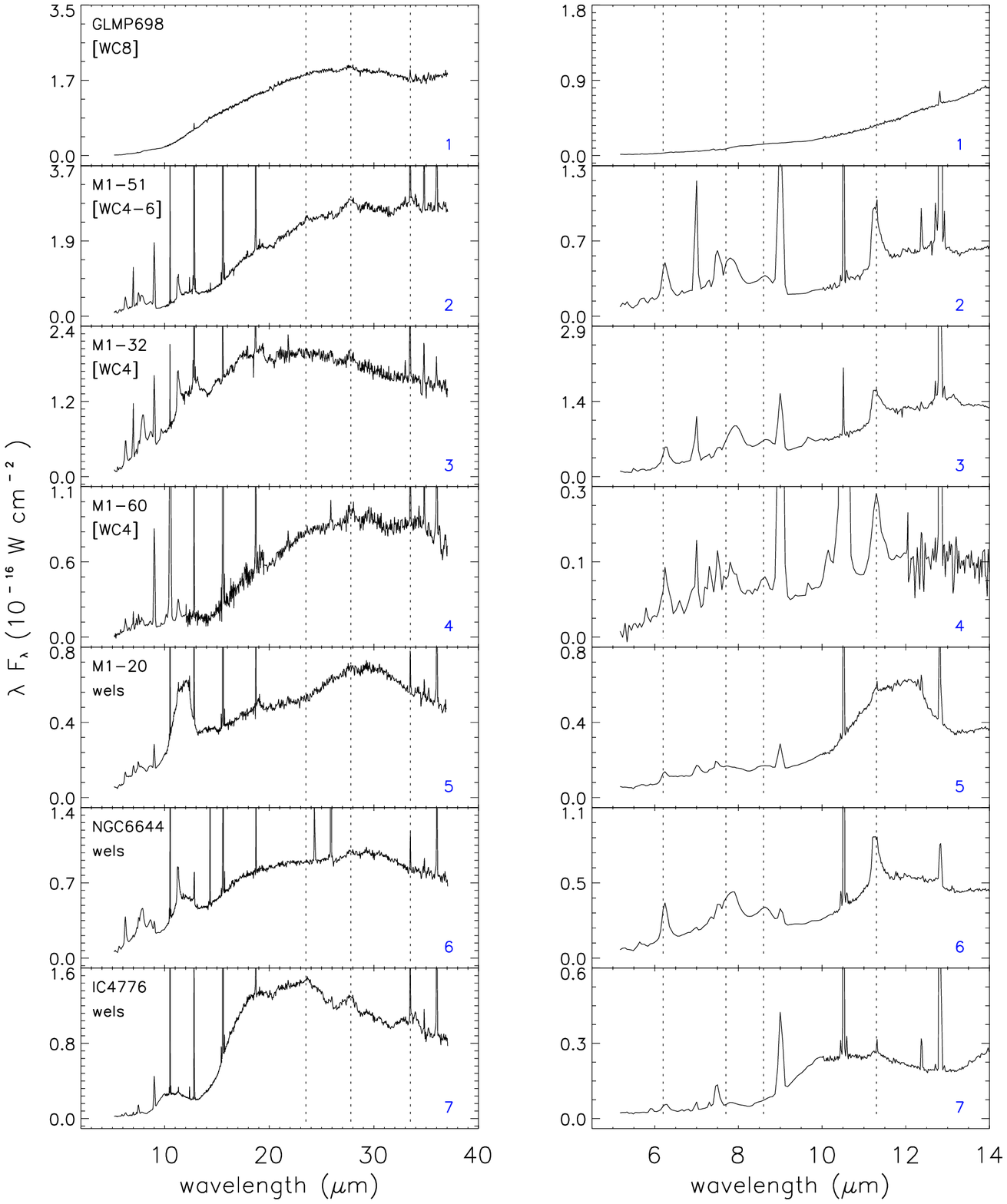}
   \caption{Spitzer/IRS spectra of PNe, which are classified here as non-Galactic
Bulge sources (see Table\,A.2). The spectral type is shown under name of PN. 
Dashed vertical lines at 23.5, 27.5 and 
33.8\,$\mu$m on the left panels mark position of crystalline silicate bands, 
while those at 6.2, ``7.7'', 8.6 and 11.3\,$\mu$m on the right panels show 
position of PAH features.}
   \label{spectranonGB1}
   \end{figure*}
   \begin{figure*}
   \centering
   \includegraphics[width=\textwidth]{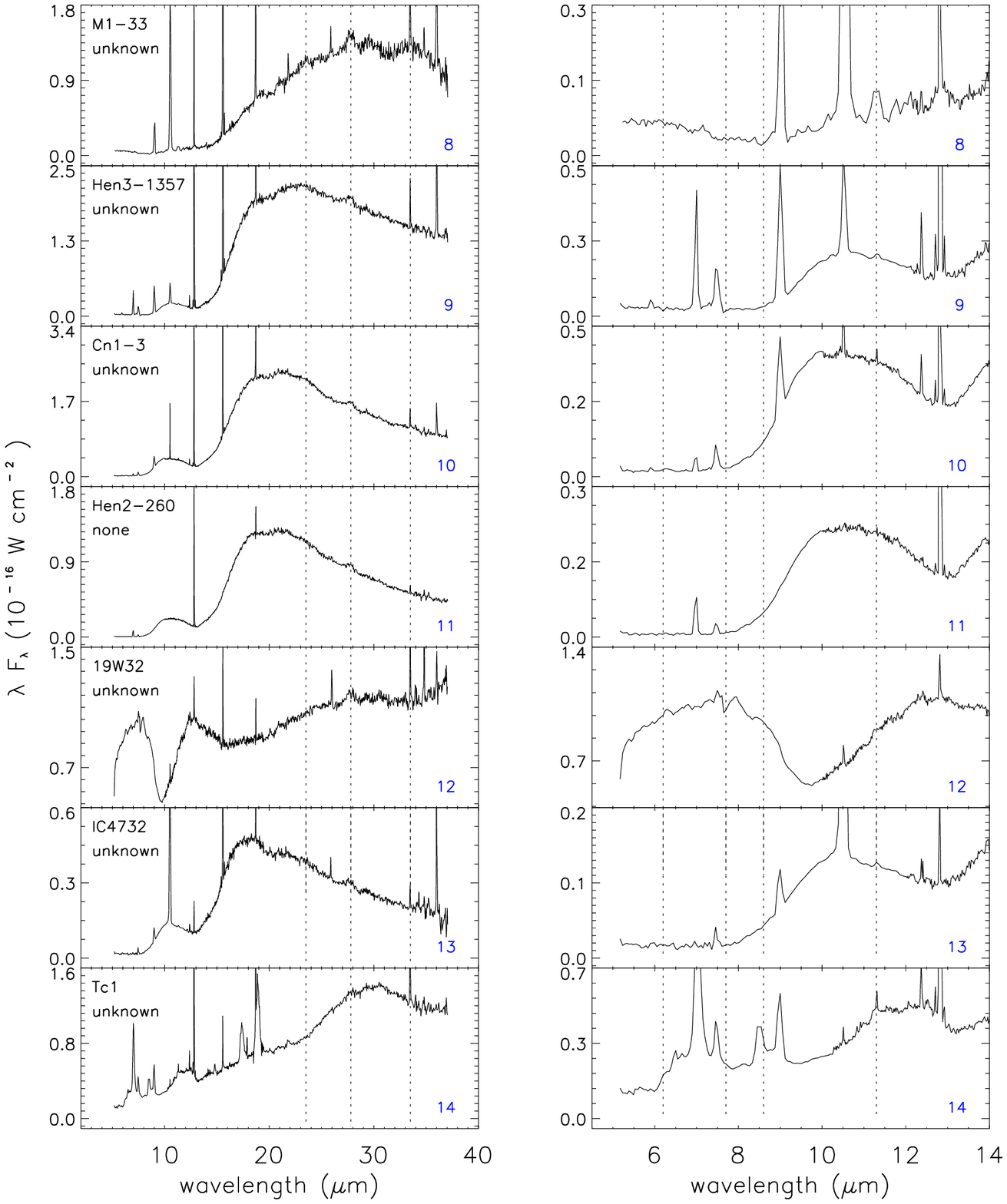}
   \caption{Spitzer/IRS spectra of PNe, which are classified here as non-Galactic
Bulge sources {\it cont.} (see Table\,A.2). The spectral type is shown under name of PN;
 none --
means neither [WC] nor {\it wels}.
Dashed vertical lines at 23.5, 27.5, and 
33.8\,$\mu$m on the left panels mark position of crystalline silicate bands, 
while those at 6.2, ``7.7'', 8.6, and 11.3\,$\mu$m on the right panels show 
position of PAH features.}
   \label{spectranonGB2}
   \end{figure*}

\end{document}